\documentclass[12pt,a4paper]{article}

\usepackage{times}
 \usepackage{graphicx}
 \usepackage{makeidx}
 \usepackage{amsmath}
 \usepackage{amssymb}
 \usepackage[mathscr]{eucal}

 \def \bfd{{\bf d}}

 \def \bfn{{\bf n}}
 
 \def \bfy{{\bf y}}

\def\bfd{{\bf d}}

\def\bfm{{\bf m}}

\def\bfy{{\bf y}}

\def\bfG{{\bf G}}

\def\bfJ{{\bf J}}

\def\calP{{\cal P}}

\def\begeq{\begin{equation}}
\def\endeq{\end{equation}}
\def\begeqarray{\begin{eqnarray}}
\def\endeqarray{\end{eqnarray}}

\begin{document} 
\title{Response to: ``A note on conditional densities, Bayes’ rule, and recent criticisms of Bayesian inference'' by Yan et al., 2026}

\author{Klaus~Mosegaard$^{1\ast}$\and
	Andrew~Curtis$^{2}$}
\date{}
\maketitle

{\small$^{1}$Niels Bohr Institute, University of Copenhagen, Copenhagen, 2200, Denmark.\and 

	\small$^{2}$ University of Edinburgh, Edinburgh, EH9 3JW, United Kingdom.} 
\vspace*{10mm}

\setcounter{page}{1}

%
%

\begin{abstract}
In a recent preprint (Mosegaard and Curtis, 2024, arXiv:2411.13570v2) we analyzed the consequences of ignoring the well-known inconsistency of classical conditional probability densities. We explained how this inconsistency, together with acausality in hierarchical methods, invalidate a variety of commonly applied Bayesian methods when
applied to problems in the physical world. Yan et al., 2026, (arXiv:2603.27038v1) published a note, in which they claim, contrary to our preprint, that there are no inconsistencies if one uses the method of conditional expectations to derive probabilities. Furthermore, they believe that there are mathematical errors in our exposition and in our use of the Bayesian framework. 
This note is a response to the claims made by Yan et al. Yan et al. do not discriminate between physical and statistical consistency. Their note addresses {\it statistical} consistency of a solution under a change of variables; this is already known to be resolved by using the theory of conditional expectations. By contrast, our preprint concerns the {\it physical} consistency of any solution under a change of mathematics used to derive that solution. It demonstrates that widely used methods to compute Bayesian posterior solutions are physically inconsistent under a change of variables. Their note does not, therefore, address the tenet of our preprint. We show herein that the theory of conditional expectations does not resolve physical inconsistency, and that Yan et al. make mathematical errors. We conclude that their claims are unfounded, and in some cases we show that their critique is meaningless. The conclusions of our preprint therefore stand.
\end{abstract}

\subsection*{Introduction}
\noindent
In a recent preprint (Mosegaard and Curtis, 2024) \cite{MosegaardCurtis2024} we analyzed the consequences of ignoring a well-known inconsistency of classical conditional probability densities, commonly referred to as the Borel Paradox. This is the apparent result that conditional probability distributions can change when we switch between different but equivalent parametrisations, if the condition is on a zero-probability event. Conditioning on zero-probability events is common in the physical sciences, in which so-called `physical laws' are imposed as exact relationships between variables in order to construct approximate models of physical reality. Bayesian inference is then often applied on those physical models to estimate conditional parameter distributions that are consistent with measured data. We explained how the Borel inconsistency, together with acausality in hierarchical methods, invalidate a variety of commonly applied Bayesian methods when
applied to problems in the physical world. 

Yan et al. \cite{Yan2026ConditionalDensities} published a note, in which they claim, contrary to our preprint, that there are no inconsistencies if one uses the method of conditional expectations to derive probabilities. Furthermore, they believe that there are mathematical errors in our exposition and in our use of the Bayesian framework. 

This note is a response to the claims made by Yan et al., which in the following is referred to as `YMMKL'.

\subsection*{On conditional expectations and consistency}
\noindent
[Response to YMMKL section 2.4]

\medskip\noindent
After having analyzed and explained the Borel Paradox, Kolmogorov, in his 1933 publication \cite{Kolmogorov1933} reestablishes mathematical consistency by proposing a new method based on conditional expectations. In a measure-theoretic framework, conditional probability densities were rigorously defined through conditional expectations with respect to an appropriately chosen sub-$\sigma$-algebra. Given a probability space with sigma algebra $\mathcal{F}$, and given a random variable $X$, the conditional expectation $\mathbb{E}[X|\mathcal{G}]$ with respect to a sub-$\sigma$-algebra $\mathcal{G}\subseteq\mathcal{F}$ provides a projection of $X$ onto the space of $\mathcal{G}$-measurable functions, ensuring consistency with the axioms of probability. The crucial step in the construction of conditional densities consists in selecting a sub-$\sigma$-algebra, typically generated by a random variable or by a measurable partition of the sample space. This selection determines the class of measurable functions that preserve the probabilistic structure of the conditioned system. Under suitable regularity conditions, the conditional density arises as the derivative of the conditional distribution with respect to an appropriate reference measure, guaranteeing uniqueness up to null sets. The method ensures compatibility with marginalization and transformation of variables. Furthermore, the sub-$\sigma$-algebra formalism clarifies the relationship between conditioning on events of probability zero and the existence of regular conditional probabilities. 

However, mathematical consistency does not induce physical consistency -- the property that every solution that combines exactly and only the same physical information, must represent exactly the same physical meaning. Physical {\em inconsistency} occurs when combining exactly the same physical information using different mathematical choices may result in physically-different solutions. 
Mathematics in itself is not physical, so it should not inject additional physical information. 

In the method using conditional expectations, {\em physical} inconsistency of the computed conditionals shows up in their dependence on the choice of sub-sigma algebra. The perhaps most well-known example is the computation of the conditional on a great circle on the unit sphere from a uniform probability density over its surface. 

If we define a geographical coordinate system in which the great circle is a meridian at longitude $0$, and we take the sub-$\sigma$-algebra generated by (measurable sets of)
meridian circles with respect to the given
North and South Poles, we obtain the conditional density $f(\theta|\phi = 0) = \cos(\theta)/2$ on the meridian, where $\theta$ is the latitude, and $\phi$ is the longitude.
If, on the other hand, we choose a sub-$\sigma$-algebra generated by measurable sets of circles on the sphere that are parallel to our meridian (that is, circles on the sphere lying in planes parallel to the plane of the meridian), we get $f(\theta|\phi = 0) = (2\pi)^{-1}$.

Much debate about the non-uniqueness of the outcome under different choices of sub-$\sigma$-algebras can be found in the mathematical-statistics literature \cite{Rescorla2015BorelKolmogorov,GyenisHoferSzaboRedei2017}, and there is generally a consensus that the BK-inconsistencies have been removed in the method of conditional expectations. The explanation is that choosing two different sub-$\sigma$-algebras corresponds to using different information about the problem. In the example with the sphere, when we consider the same great circle using two different sub-$\sigma$-algebras, and hence calculate the conditional probability distribution using different versions of the conditional expectations, then we do {\em not} ``calculate conditional probabilities in different coordinate systems", but rather, we {\em calculate conditional probabilities with respect to different conditioning sub-$\sigma$-algebras} \cite{GyenisHoferSzaboRedei2017}.

Seen from a purely logical/mathematical viewpoint, this is indeed a satisfactory resolution of the BK-inconsistency. In physical sciences, however, these arguments are invalid. Invariance of any statement about nature under changes of mathematical description is essential: the physical properties of reality, and our information about it, do not depend on its mathematical representation 
\cite{Einstein1916,Jeffreys1946,Norton1993,Thorne2017}. If we accept that any parameter of a physical system and its probability distribution represent a (possibly observer-dependent) physical property, the {\em principle of covariance} imposes invariance under changes of mathematical description: Bayesian inversion schemes that do not satisfy this principle may lead to physically unacceptable solutions. 

So, in the example of the sphere, why is it that we, from a physical standpoint, must require that the conditional on a great cicle must be independent of specification of a coordinate system or a sub-$\sigma$-algebra? The answer lies in an understanding of what a physical parameters is: A physical parameter is not just a number, but a {\em potentially observable number}, usually equipped with a physical unit. Examples are data and unknown physical parameters in Bayesian inversion. Consider an example with meteorites impacting on a spherical planet, where we initially know that the distribution of impact points is uniform over the surface. Impact points are physically measurable without use of coordinates, for instance by measuring the distance to 3 fixed points on the planetary surface using a yard stick (a rigid body, defined as a standard meter). 
As the (measurable) time tends to infinity, the normalized, measurable density of physical impact points -- the number of impact points in a unit area defined by the yard stick -- tends to the probability density over the surface.
If we are now given the additional information from an observer that (the centers of mass of) a particular family of meteorites, sharing similar orbital elements, will hit on a given great circle, but with no information about the impact point on the great circle, we can ask: What is the conditional impact probability density on the great circle of this family? Physically, the great circle is a geometrical object, completely defined by its points on the surface of the planet, and, as time tends to infinity, the measurable density of physical impact points (impacts per unit length) tends to the conditional probability density over the meridian. This density is constant, it is unique, and it is independent of any choice of parameterization or sub-$\sigma$-algebra.

No matter which (sensible) parameterization or sub-$\sigma$-algebra we choose in order to carry out our calculations, the results should be the same. Mathematical choices are not based on physical measurements, so if they change our results then we have added spurious (unphysical) information to our inference problem.

The conclusion is that the method of conditional expectations removes the inconsistencies seen from a purely logical/mathematical viewpoint, but fails to live up to the principle that physical predictions must be fundamentally independent of arbitrary choices, such as the mathematical framework, coordinate system, or gauge chosen to describe them, because physical laws are generalizations of observable, objective, immutable reality.

A corollary is that either (a) we require that there is a mathematical description of the problem that injects zero additional physical information; or (b) that there exists a unique sub-$\sigma$-algebra that is defined unambiguously and required by the physics of the problem. Neither of these appears to be true.

\subsection{Any evidence can be obtained by choosing an
appropriate re-parameterization in the data space}
\noindent
[Response to YMMKL section 3.1]

\medskip\noindent
Their critique reads: {\em MC’s Appendix G demonstrates that different models can yield different results. ...Obviously, different models can yield different results, if those models are not equivalent.} 

In terms of critique, this comment by YMMKL is meaningless: it adds nothing. After an apparently hasty reading, YMMKL fail to acknowledge that the demonstration in G.0.1 should be combined with the subsequent demonstration G.0.2 to show that, for an overdetermined inverse problem with parameter space $[0,1]^N$, any evidence can be obtained by choosing an appropriate re-parameterization in the data space.

\subsection{Inconsistency of our simple tomographic example}
\noindent
[Response to YMMKL section 3.2]

\medskip\noindent
Their critique reads: {\em MC’s “simple tomographic example” (in their Appendix A) is invalid: conditioning on a zero-probability event is not allowed} 

\smallskip\noindent 
If one intends to criticize or refute the messages in our text publicly, the first requirement must be to read the text thoroughly. In the last paragraph of the introductory text (before the section {\em The basic assumptions}) it is clearly stated that ``The demonstrations are based on {\it reductio ad absurdum}, that is, analytical examples where the methods in question, applied to physical inference problems, are shown to be mathematically self-contradictory or in other ways unacceptable.''. In other words, our purpose is indeed to show that our “simple tomographic example” is inconsistent. Our reason for showing the example is that this demonstrably inconsistent method is widely used by practitioners; to criticise the method is merely to agree with our point.

\subsection{Setup of our examples concerning acausality}
\noindent
[Response to YMMKL section 3.2]

\medskip\noindent
Their critique reads: {\em MC’s examples concerning acausality (in their Appendix F) are invalid: their models have not been set up correctly}

\smallskip\noindent 
After having critizised our setup of the problem they do the calculations themselves (in the `correct' way), and they arrive at exactly the same result!
To see this, note that their $y$ is our $\bfd_{obs}$, and their $km$ is our $\bfd$. 

In general, they seem to have read our section on `Formulation 2' very superficially. We describe the notation in that section, and even if they, for some reason, dislike our interpretation of $\bfd$, they should be able to see that `Formulation 2' amounts to defining a new density
\begin{equation*}
p_x(x) \equiv p_{n}(\bfd_{obs}-x) \ ,
\end{equation*}
where $p_{n}$ is the noise distribution, and $\bfd_{obs}$ is the observed data. From this it follows that the likelihood becomes $p_{g(\bfm)}(g(\bfm))$, which is $p_{d}(\bfd)$ conditioned on $\bfd = g(\bfm)$. For this reason, it is not surprising that using $p_{d}(\bfd)$ as likelihood gives the same results as using $p_{n}(\bfd_{obs}-g(\bfm))$ as likelihood. Those expressions are equal.

Our conclusion therefore holds. The computed `hyperparameters' depend on the forward function (through $k$). The values of the hyperparameters are therefore posterior values, and this means that the distributions they define are posteriors, not priors. Note, that there is no way to compute the hyperparameters without involving the data and the forward function: this is where the information is coming from!

\subsection{Consequences of change of variables in data space}
\noindent
[Response to YMMKL section 4 and 4.1]

\medskip\noindent
Their critique reads: {\em MC’s examples involving MAP estimation, empirical Bayes and Bayes factors (in their Appendices B–D) are invalid: reparametrisation is performed incorrectly}. Plus some additional criticisms.

\subsubsection{Response to critique of our Appendix B}
First, in our Appendix B, the data are not mentioned at all, and a completely regular Jacobian transformation is applied in the parameter space, and only there. For this reason, YMMKL's comment that we ``have evaluated the Jacobian not at the observed data y but at the value predicted by the forward
model g(m)'' is meaningless. That we ``have gone wrong'' is a surprising comment, since the inconsistency of MAP estimates is well-known in mathematical statistics.

\medskip\noindent
To respond to YMMKL's comments on Appendices B–D, we need to go deeper into Formulation 2:

\smallskip\noindent 
\subsubsection{A note on our Formulation 2}
Formulation 2 is a direct model of physical measurement and subsequent inference. In detail, it involves the following steps:
\begin{enumerate}
\item
Establish/collect prior information (defined as any information that is independent of the physical relation between data and model parameters):
\begin{itemize}
\item Prior information in the parameter space, given as a prior distribution $p(\bfm)$.
\item Prior information in the data space, given as the distribution $p(\bfd)$ of the (unknown) {\em noise-free} data.
\end{itemize}
Notice that `data' and `model parameters' are both physical parameters, with the exception that data are observable, and model parameters are non-observable. Otherwise there is no fundamental difference between the two kinds of parameters. This means that, in a Bayesian setting, we will in general have prior information about both data and model parameters. 
For the data, the prior expresses the results of a measurement: typically, it is the noise distribution, shifted to be centered at the observations.   

\item
Form the joint prior $p_{d,m} (\bfd,\bfm)$. If data and model parameters are independent, $p_{d,m} (\bfd,\bfm) = p(\bfd)p(\bfm)$.

\item
Finally, introduce the additional information that there is a relation $h(\bfd,\bfm) = 0$ between $\bfd$ and $\bfm$. Typically, $\bfd$ is a function of $\bfm$, say, $\bfd = g(\bfm)$.

\item
Solve the problem by conditioning $p_{d,m} (\bfd,\bfm)$ on $\bfd = g(\bfm)$.


\end{enumerate}
Incidentally, if you do not subscribe to this view, but prefer the classical setup, `Formulation 2' can be seen as a simple and harmless substitution of $p_{n}(\bfd_{obs}-x)$ with $p_x(x)$.

\subsubsection{Response to critique of our Appendices C and D}

Our Formulation 2 gives the same results as Formulation 1. YMMKL forget that, when they transform the data, they must also transform the forward function $g(\bfm)$ -- a mathematical error.
Consider a standard derivation of the likelihood in our {\em Appendix C.2 Case 2: Transformed data with $d_1 \rightarrow \tan(d_1)$}: 

\label{Appendix: Likelihood-Formulation-1}

\bigskip\noindent
Let us reconsider the linear inverse problem $\bfd^{obs} = \bfG \bfm + \hat{\bfn}$: 
\begin{equation}
\begin{pmatrix}
d_1\\
d_2\\
d_3
\end{pmatrix} =
\begin{Bmatrix}
0 & a\\
b & 0\\
c & 0
\end{Bmatrix} 
\begin{pmatrix}
m_1\\
m_2
\end{pmatrix}
+ \begin{pmatrix}
\hat{n}_1\\
\hat{n}_2\\
\hat{n}_3
\end{pmatrix}
\label{lin-probl-2}
\end{equation}
where $\bfd^{obs} = (d_1, d_2, d_3)^T$ is the observed data, and $\bfm = (m_1, m_2)^T$ is the unknown parameter vector. The noise $\hat{\bfn} = (\hat{n}_1, \hat{n}_2, \hat{n}_3)^T$ has the distribution
\begin{equation}
   p_d(\hat{\bfn}) = 
   \begin{cases}
       \frac{1}{(2\sigma)^3} & {\rm for} \ \hat{\bfn} 
             \in [-\sigma,\sigma]^3 \\
       0 & \text{otherwise} \ .
   \end{cases} 
\label{eq: data-distrib-cart}
\end{equation}
Since this distribution is symmetric around the origin, the sign-reversed noise (to be used below) $\bfn = -\hat{\bfn}$ has the same distribution:
\begin{equation}
   p_d(\bfn) = 
   \begin{cases}
       \frac{1}{(2\sigma)^3} & {\rm for} \ \bfn 
             \in [-\sigma,\sigma]^3 \\
       0 & \text{otherwise} \ .      
   \end{cases} 
\label{eq: data-distrib-cart}
\end{equation}
Data and uncertainties are now reparameterized via the transformation
\begin{equation}
\begin{pmatrix}
y_1\\
y_2\\
y_3
\end{pmatrix} =
T_1
\begin{pmatrix}
d_1\\
d_2\\
d_3
\end{pmatrix} =
\begin{pmatrix}
\tan(d_1)\\
d_2\\
d_3
\end{pmatrix} \ ,
\end{equation}
with absolute determinant of the inverse Jacobian
\begin{equation}
|\det(\bfJ_S)| = \left| \frac{\partial(d_1, d_2, d_3)}{\partial(y_1, y_2, y_3)} \right| = \frac{1}{y_1^{2}+1} \ .
\label{Jac-arctan}
\end{equation}
The transformed forward function is
\begin{equation}
    g_y(\bfm) = 
    \begin{pmatrix}
        \tan(a m_2)\\
        bm_1\\
        cm_1
   \end{pmatrix} \ .
\end{equation}
Substituting $y_1 = y_1^{obs}+\hat{n}_1 = y_1^{obs}-n_1$ into equation \ref{Jac-arctan}, the Jacobian transformation provides the distribution of noise on the transformed data $\bfy$:
{\footnotesize
\begin{equation*}
    p_n^{y} (\bfn) = 
       \begin{cases}
       \frac{1}{(2\sigma)^3}\frac{1}{(y_1^{obs}-n_1)^{2}+1} & {\rm for} \ \bfn \in 
          [\tan(d_1-\sigma)-\tan(d_1),\tan(d_1+\sigma)-\tan(d_1)] \times [-\sigma,\sigma]^2 \\
       0 & \text{otherwise}          
   \end{cases}
\end{equation*}
}where we have carried out a Jacobian transformation using equation (\ref{Jac-arctan}). 

The standard likelihood expression for the inverse problem with $\bfy$ as data is (see our {\em Formulation 1} or, e.g., Calvetti and Somersalo, 2017) \cite{CalvettiSomersalo2018}:
\begin{align*}
    L_y(\bfm) &= p_n^{y} (\bfy - g_y(\bfm)) = 
        p_n^{y} \left(
        \begin{pmatrix}
            y_1 - \tan(a m_2)\\
            y_2 - bm_1\\
            y_3 - cm_1
        \end{pmatrix} \right) \\
        &= 
         \begin{cases}
            \frac{1}{(2\sigma)^3}\frac{1}{\tan(a m_2)^{2}+1} & {\rm for} \ \bfm \in \calP(\sigma)  \\
            0 & \text{otherwise}
         \end{cases} \\
         &= 
         \begin{cases}
            \frac{1}{(2\sigma)^3}\cos^2(a m_2) & {\rm for} \ \bfm \in \calP(\sigma)  \\
            0 & \text{otherwise}
         \end{cases}
\end{align*}
where $\calP(\sigma)$ is the domain in the parameter space with non-zero likelihoods. In contrast to the expectations of YMMKL, the above result is identical to the solution that we obtained in Appendix C.2 Case 2, with formulation 2 of Bayes Formula.

\subsection*{Conclusions}

Yan et al. (2026) published a note, in which they claim, contrary to our preprint (Mosegaard and Curtis, 2024) \cite{MosegaardCurtis2024}, that there are no inconsistencies if one uses the method of conditional expectations to derive probabilities. Furthermore, they believe that there are mathematical errors in our exposition and in our use of the Bayesian framework. This note is a response to these claims. 

Yan et al. do not discriminate between physical and statistical consistency. Their note merely addresses the {\it statistical} consistency of a solution under a change of variables; this is already known to be resolved by using the theory of conditional expectations. By contrast, our preprint concerns the {\it physical} consistency of any solution under a change of mathematics used to derive that solution. The preprint demonstrates that widely used methods to compute Bayesian posterior solutions are physically inconsistent under a change of variables. The note from Yan et al. does not, therefore, address the tenet of our preprint. We show herein that the theory of conditional expectations does not resolve physical inconsistency, and that Yan et al. make mathematical errors. We conclude that their claims are unfounded, and in some cases we show that their critique is meaningless.

\clearpage

\subsubsection{}

\bibliographystyle{sciencemag}
\bibliography{BayesModSelectReferences}{}

@article{CalvettiSomersalo2018,
  author  = {Calvetti, Daniela and Somersalo, Erkki},
  title   = {Inverse problems: From regularization to Bayesian inference},
  journal = {WIREs Computational Statistics},
  year    = {2018},
  volume  = {10},
  number  = {1},
  pages   = {e1427},
  doi     = {10.1002/wics.1427},
  url     = {https://doi.org/10.1002/wics.1427}
}

@article{Einstein1916,
  author    = {Albert Einstein},
  title     = {The foundation of the general theory of relativity},
  journal   = {Annalen der Physik},
  year      = {1916},
  volume    = {49},
  number    = {7},
  pages     = {769--822}
}

@article{GyenisHoferSzaboRedei2017,
  author    = {Zolt{\'a}n Gyenis and G{\'a}bor Hofer-Szab{\'o} and Mikl{\'o}s R{\'e}dei},
  title     = {Conditioning using conditional expectations: The Borel--Kolmogorov Paradox},
  journal   = {Synthese},
  year      = {2017},
  volume    = {194},
  number    = {7},
  pages     = {2595--2630},
  doi       = {10.1007/s11229-016-1040-2}
}

@article{Jeffreys1946,
  author    = {Harold Jeffreys},
  title     = {An invariant form for the prior probability in estimation problems},
  journal   = {Proceedings of the Royal Society of London. Series A. Mathematical and Physical Sciences},
  year      = {1946},
  volume    = {186},
  number    = {1007},
  pages     = {453--461}
}

@book{Kolmogorov1933,
  author    = {Andrey N. Kolmogorov},
  title     = {Foundations of the Theory of Probability},
  year      = {1933/1956},
  address   = {New York},
  publisher = {Chelsea},
  note      = {2nd English ed., Trans. N. Morrison}
}

@article{MosegaardCurtis2024,
  author       = {Mosegaard, Klaus and Curtis, Andrew},
  title        = {Inconsistency and acausality in Bayesian inference for physical problems},
  journal      = {arXiv preprint},
  year         = {2024},
  eprint       = {2411.13570},
  archivePrefix= {arXiv},
  primaryClass = {stat.ME},
  note         = {Version 2, accessed 29 January 2026},
  url          = {https://arxiv.org/abs/2411.13570}
}

@article{Norton1993,
  author    = {John D. Norton},
  title     = {General covariance and the foundations of general relativity: eight decades of dispute},
  journal   = {Reports on Progress in Physics},
  year      = {1993},
  volume    = {56},
  number    = {7},
  pages     = {791--858},
  doi       = {10.1088/0034-4885/56/7/001},
  s2cid     = {250902085}
}

@article{Rescorla2015BorelKolmogorov,
  author  = {Michael Rescorla},
  title   = {Some Epistemological Ramifications of the Borel-Kolmogorov Paradox},
  journal = {Synthese},
  year    = {2015},
  volume  = {192},
  number  = {3},
  pages   = {735--764},
  doi     = {10.1007/s11229-014-0560-1}
}

@book{Thorne2017,
  author    = {Kip S. Thorne and Roger D. Blandford},
  title     = {Modern Classical Physics},
  publisher = {Princeton University Press},
  year      = {2017}
}

@article{Yan2026ConditionalDensities,
  author       = {Yan, Alex and Mills, Cathal and Marignier, Augustin and Kim, Younjung and Lambert, Ben},
  title        = {A note on conditional densities, Bayes' rule, and recent criticisms of Bayesian inference},
  journal      = {arXiv preprint arXiv:2603.27038},
  year         = {2026},
  eprint       = {2603.27038},
  archivePrefix= {arXiv},
  primaryClass = {stat.ME},
  doi          = {10.48550/arXiv.2603.27038},
  url          = {https://arxiv.org/abs/2603.27038}
}

%
%
%
%
%

\renewcommand{\labelenumi}{\theenumi.}
\clearpage

\end{document}